\documentclass{jps-cp}
\usepackage{txfonts} 

\usepackage{url}
\usepackage{epsfig}

\newcommand{\mi}{\mathrm{i}} 

\newcommand{\dfd}[3]{\hspace{-0.4em}\ensuremath{\frac{\mathrm{d}^{#1}#3}{(2\pi)^{#2}}}\,}

\newcommand{\eqn}[1]{Eq.~(\ref{#1})}
\newcommand{\fig}[1]{Figure~\ref{#1}}
\newcommand{\tab}[1]{Table~\ref{#1}}

\def\Kbar{\overline{K}}
\def\K0bar{\overline{K^0}}

\title{
\vspace{-25mm}
\begin{flushright}
\texttt{LFTC-19-4/42}
\end{flushright}
\vspace{25mm}
$\Phi$ and $J/\Psi$ Mesons in Cold Nuclear Matter}

\author{J.~J.~\textsc{Cobos-Mart\'{\i}nez}$^{1}$, K.~\textsc{Tsushima}$^{2}$,
  G.~\textsc{Krein}$^{3}$, and A.~ W.~\textsc{Thomas}$^{4}$}

\inst{$^{1}$ CONACyT--Departamento de F\'{\i}sica, Centro de
Investigaci\'on y de Estudios Avanzados del Instituto Polit\'ecnico
Nacional, Apartado Postal 14-740, 07000, Ciudad de M\'exico, M\'exico. \\
$^{2}$Laborat\'orio de F\'{\i}sica Te\'orica e Computacional-LFTC,
Universidade Cruzeiro do Sul, 01506-000, S\~ao Paulo, SP, Brazil. \\
$^{3}$Instituto de F\'{\i}sica Te\'orica, Universidade Estadual
Paulista, Rua Dr. Bento Teobaldo Ferraz, 271-Bloco II, 01140-070, S\~ao
Paulo, SP, Brazil. \\
$^{4}$ ARC Centre of Excellence for Particle Physics at the Terascale
and CSSM, Department of Physics, University of Adelaide, Adelaide SA
5005, Australia.}

\email{jcobos@fis.cinvestav.mx}

\recdate{To be inserted}

\abst{We present $\Phi$- and $J/\Psi$--nuclear bound state energies and
absorption widths for some selected nuclei, using potentials in the
local density approximation computed from an effective Lagrangian approach
combined with the quark-meson coupling model.
Our results suggest that these mesons should form bound states with all the
nuclei considered provided that these mesons are produced in nearly recoilless
kinematics.}

\kword{vector mesons, nuclear matter, vector meson mass shift, 
nuclear bound states}

\begin{document}

\maketitle

\section{Introduction}
\vspace{-1.25mm}

The properties of vector mesons at finite baryon density, such as its mass
and decay width, have attracted considerable experimental and theoretical
interest over the last few decades~\cite{vectormesonsinnuclmatt}, in part 
due to their potential to carry information on the partial restoration of
chiral symmetry and the possible role of QCD van der Waals forces in the
binding of quarkonia to nuclei~\cite{vanderwaals}.
However, an experimentally unified consensus has not yet been reached for the
$\phi$ meson~\cite{phipptiesnuclmatt} and further studies need to be done
~\cite{JPARCE29Proposal,JLabphiProposal}.
The study of, for example, the $\phi$--nucleus bound
states~\cite{JPARCE29Proposal, JLabphiProposal} is expected to provide
information on the $\phi$ properties at finite density, since a downward 
mass shift of the $\phi$ in a nucleus is directly connected with the 
existence of an attractive potential between the $\phi$ and the nucleus 
where it has been produced.
Various authors predict a small downward shift of 
the in-medium $\phi$ mass and a large broadening of its decay width~\cite{phipptiestheory} at normal nuclear matter density.
In Ref.~\cite{Cobos-Martinez:2017vtr} we computed the $\phi$ mass shift and
decay width in nuclear matter by evaluating the $K\overline{K}$ loop
contribution to the $\phi$ self-energy, with the in-medium $K$ and
$\overline{K}$ masses calculated using the quark-meson coupling
(QMC) model~\cite{Tsushima:1997df}. This study was extended
in Ref.~\cite{Cobos-Martinez:2017woo} by computing the $\phi$--nucleus bound
state energies and absorption with complex potentials. The Results for
$^{197}\text{Au}$ nucleus are presented for the first time. Furthermore, we also
update results for the $J/\Psi$ vector meson, adding also the $^{197}\text{Au}$
nucleus for the first time.

\section{$\Phi$-meson in nuclear matter and $\Phi$-meson--nucleus bound states}

\begin{figure}
\begin{center}
\begin{tabular}{ccc}
\includegraphics[scale=0.203]{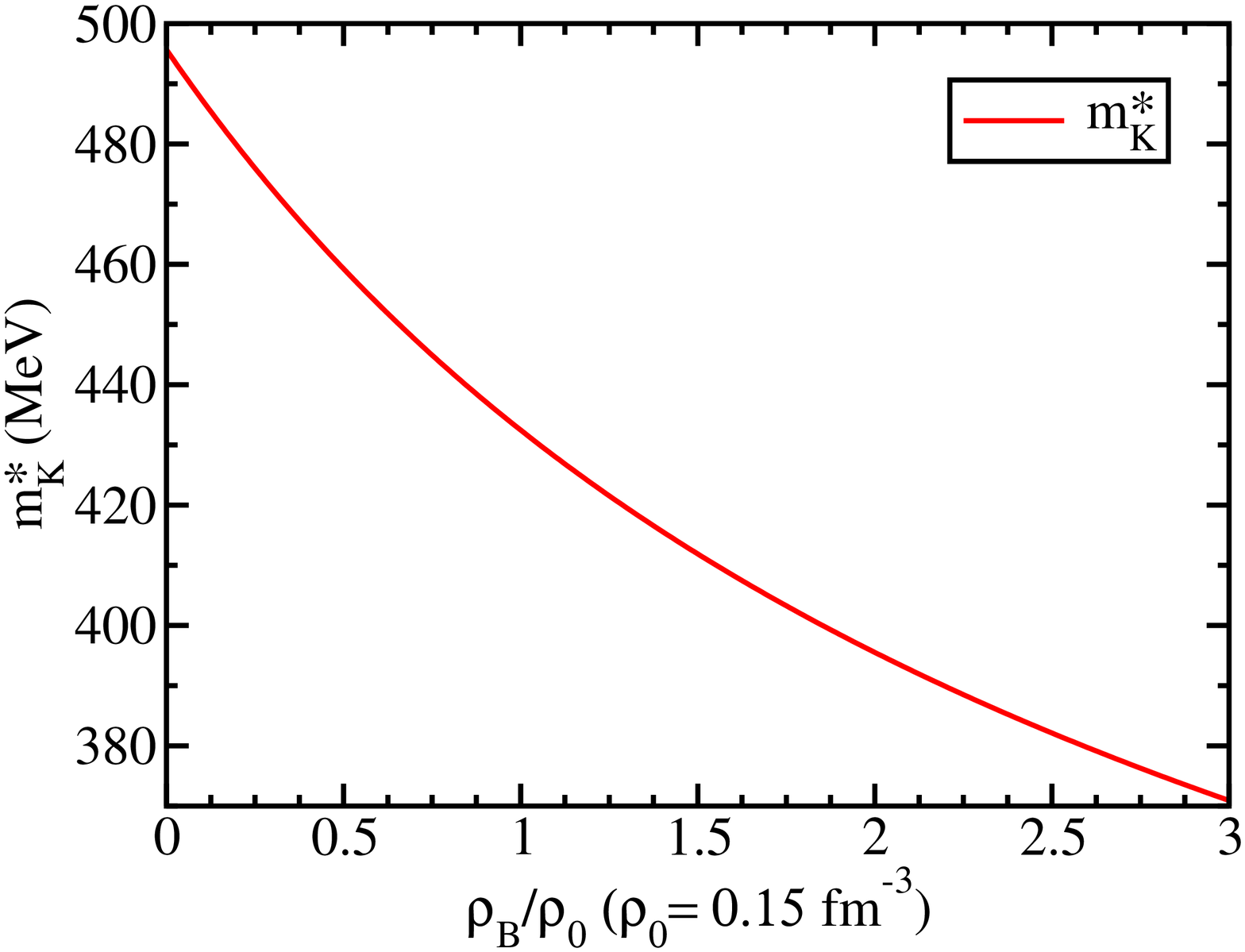} &
\includegraphics[scale=0.177]{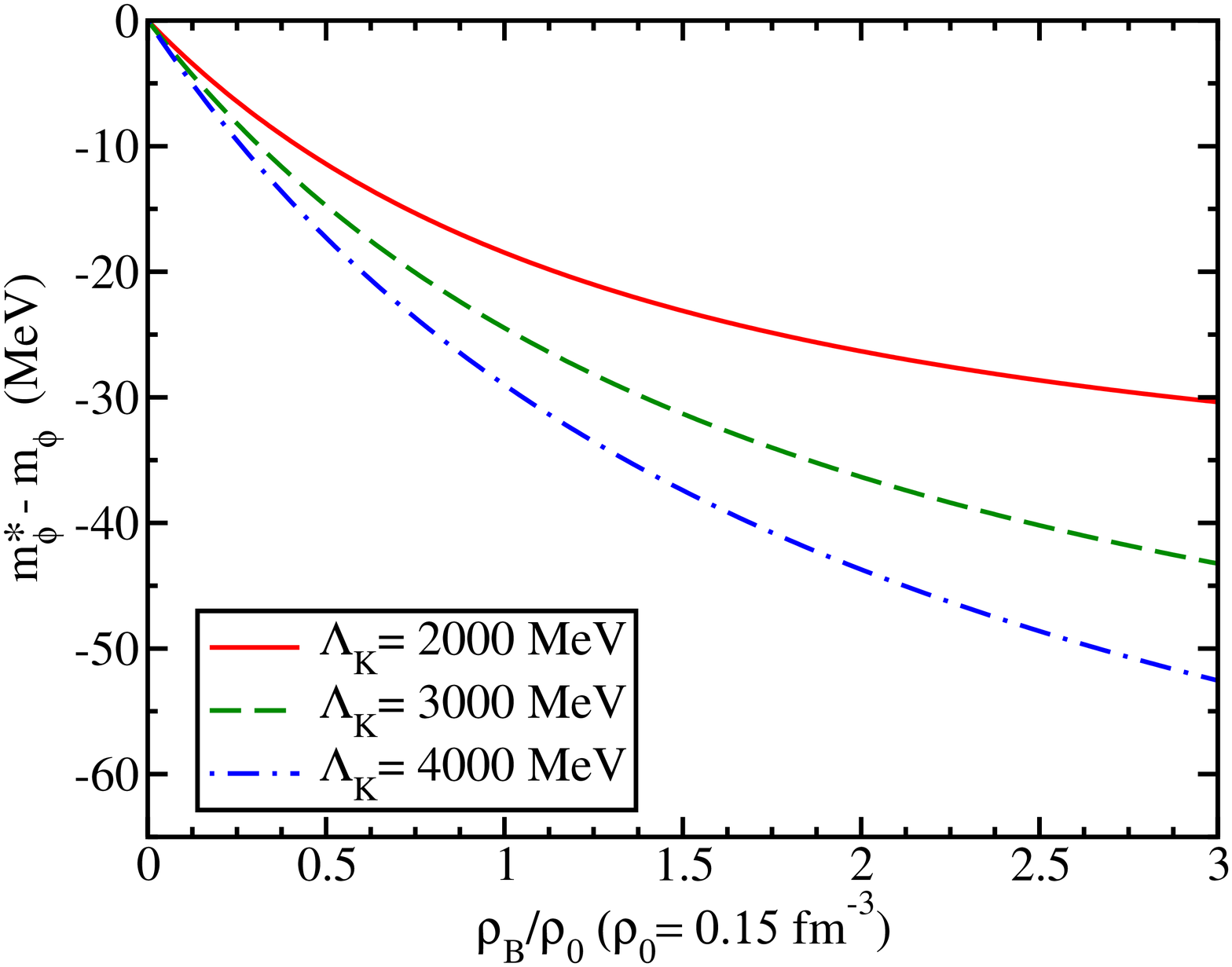} &
 \includegraphics[scale=0.177]{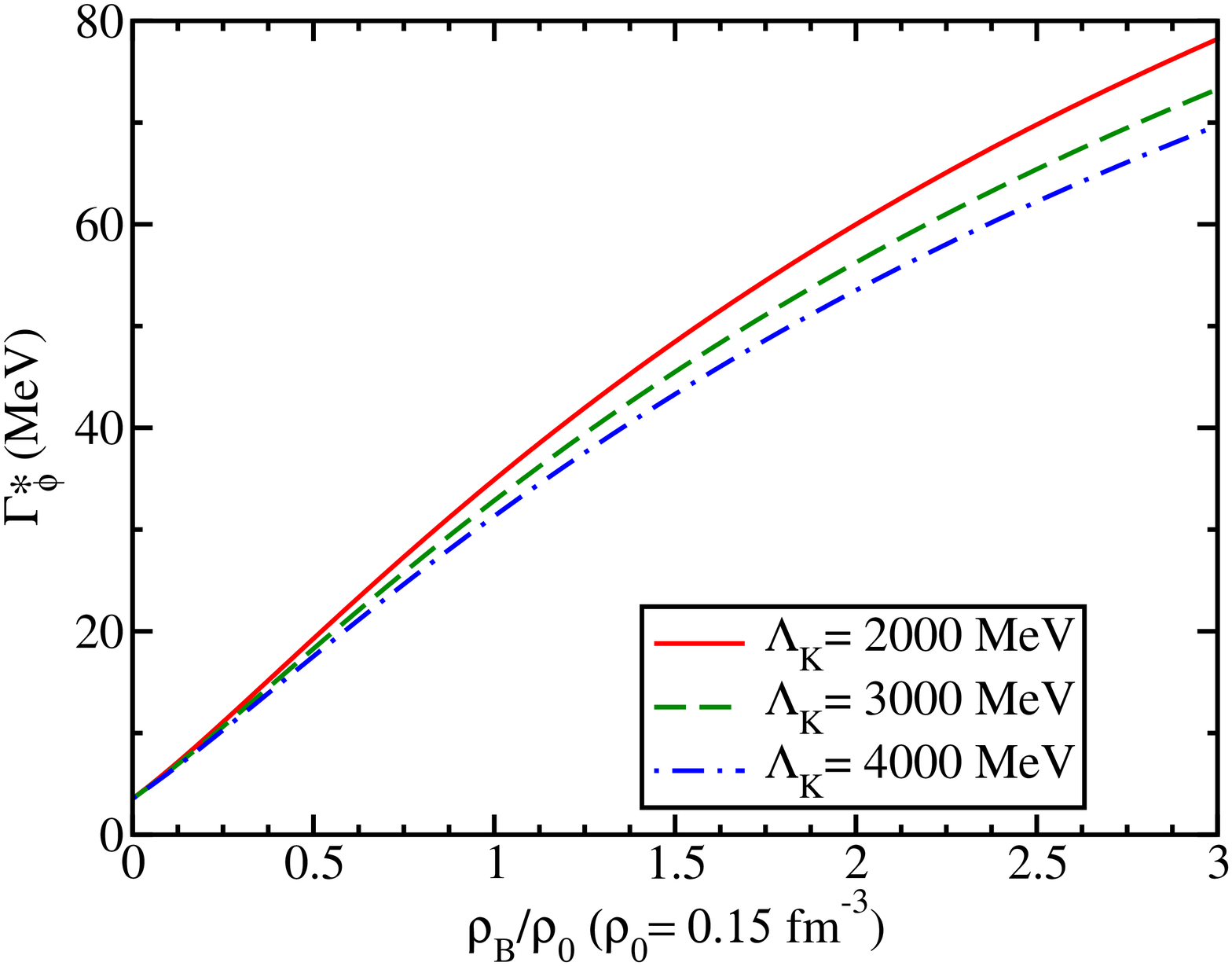}
\end{tabular}
\caption{\label{fig:nuclmatt} Left panel: In-medium kaon mass; center and right panels: $\phi$ mass shift and decay width.}
\end{center}
\end{figure}

We compute the $\phi$ self-energy $\Pi_{\phi}$ in vacuum and in nuclear matter~\cite{Cobos-Martinez:2017vtr} using an effective Lagrangian approach,
considering only the $\phi K\overline{K}$ vertex~\cite{Cobos-Martinez:2017vtr}, 
since we expect that a large fraction of the density dependence of $\Pi_{\phi}$
arises from  the in-medium modification of the $K\Kbar$ intermediate state,
\begin{equation}
\label{eqn:Lpkk}
\mathcal{L}_{\phi K\overline{K}}=\mi g_{\phi}\phi^{\mu}
[\Kbar(\partial_{\mu}K)-(\partial_{\mu}\Kbar)K],
\end{equation}
\noindent where $K$ and $\Kbar$ are isospin doublets and $\phi^{\mu}$ is
the $\phi$ meson vector field.
The contribution from \eqn{eqn:Lpkk} to $\Pi_{\phi}(p)$ is given by 
\begin{equation}
\label{eqn:phise}
\mi\Pi_{\phi}(p)=-(8/3)g_{\phi}^{2}\int\dfd{4}{4}{q}\vec{q}^{\,2}
D_{K}(q)D_{K}(q-p),
\end{equation}
\noindent where $D_{K}(q)=1/(q^{2}-m_{K}^{2}+\mi\epsilon)$ is the
kaon propagator;  $p=(p^{0}=m_{\phi},\vec{0})$ for a $\phi$ at rest, 
$m_{\phi}$ its mass; $m_{K} (=m_{\Kbar})$ the kaon mass; and
$g_{\phi}= 4.539$~\cite{Cobos-Martinez:2017vtr} the coupling constant. 
The integral in \eqn{eqn:phise} is divergent and will be regulated using a
dipole form factor, with cutoff parameter
$\Lambda_{K}$~\cite{Cobos-Martinez:2017vtr}. The dependence of our results 
on the value of $\Lambda_{K}$ is studied below.
The mass and decay width of the $\phi$ in vacuum ($m_{\phi}$ and
$\Gamma_{\phi}$), as well as in nuclear matter ($m_{\phi}^{*}$ and
$\Gamma_{\phi}^{*}$), are determined~\cite{Cobos-Martinez:2017vtr} from
\begin{equation}
\label{eqn:phippties} 
m_{\phi}^{2}=(m_{\phi}^{0})^{2}+\operatorname{Re}\Pi_{\phi}(m_{\phi}^{2}), 
\quad \Gamma_{\phi}=-(1/m_{\phi})\operatorname{Im}\Pi_{\phi}(m_{\phi}^{2}).
\end{equation}
The density dependence of the $\phi$ mass and decay width is driven by the
interactions the $K\Kbar$ intermediate state with the nuclear medium, which
we calculate in the QMC model~\cite{Tsushima:1997df,Saito:2005rv}.
In \fig{fig:nuclmatt} (left panel) we present the in-medium kaon
Lorentz scalar mass as a function of the baryon density. At normal nuclear
matter density $\rho_{0}= 0.15$ fm$^{-3}$ $m_{K}^{*}$ has decreased by 13\%.
In \fig{fig:nuclmatt} we present the $\phi$ mass shift (center panel)
and decay width (right panel) as a function of the nuclear matter density,
$\rho_{B}$, for three values of $\Lambda_{K}$. 
For the largest value of $\rho_{B}$, the downward mass shift turns out to be
a few percent at most for all $\Lambda_{K}$. On the other hand,
$\Gamma_{\phi}^{*}$ depends strongly on the nuclear density, increasing 
by up to a factor of $\approx 20$ for the largest value of $\rho_{B}$.
These results open the experimental possibility to study the binding and
absorption of the $\phi$ in nuclei. 


We now investigate the situation where the $\phi$ meson is ``placed" inside
a nucleus~\cite{Cobos-Martinez:2017woo}.
The nuclear density distributions for all nuclei but $^{4}$He are obtained
in the QMC model~\cite{Saito:1996sf}. For $^{4}$He we use
Ref.~\cite{Saito:1997ae}. Using a local density approximation the $\phi$--nucleus potentials for a nucleus $A$ is given by
\begin{equation}
\label{eqn:Vcomplex}
V_{\phi A}(r)= U_{\phi}(r)-(\mi/2)W_{\phi}(r),
\end{equation}
\noindent where $r$ is the distance from the center of the nucleus and 
$U_{\phi}(r)=m^{*}_{\phi}(\rho_{B}^{A}(r))-m_\phi$ and $W_{\phi}(r)=\Gamma_{\phi}(\rho_{B}^{A}(r))$ are, respectively, the $\phi$
mass shift and decay width inside nucleus $A$, with $\rho_{B}^{A}(r)$ 
the baryon density distribution of nucleus $A$.
\begin{figure}
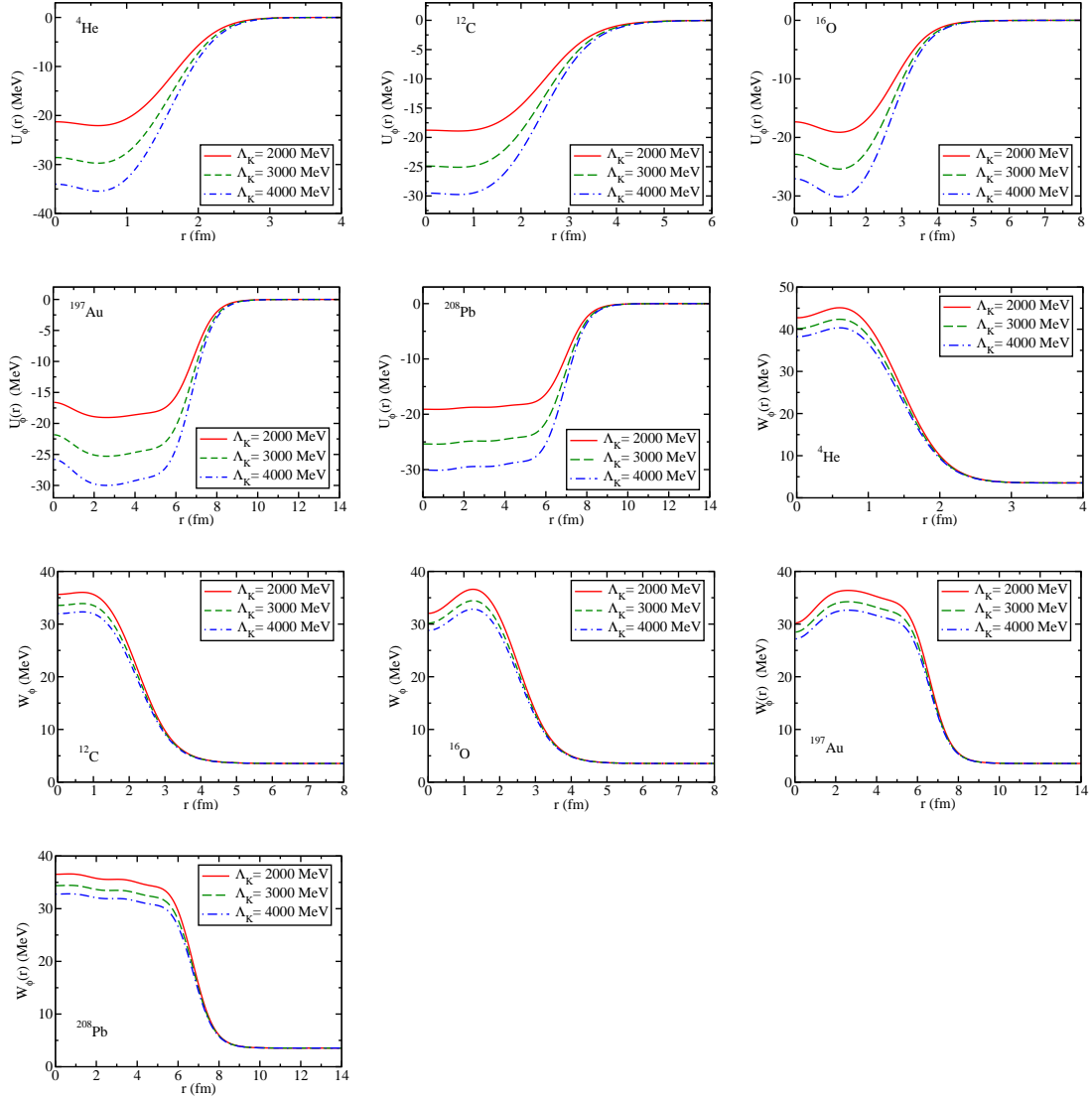

  \centering
{\renewcommand{\arraystretch}{4.00}%
\begin{tabular}{ccc}
  \includegraphics[scale=0.187]{Vphi_He4.eps} &
  \includegraphics[scale=0.187]{Vphi_C12.eps} &
  \includegraphics[scale=0.187]{Vphi_O16.eps} \\
  \includegraphics[scale=0.187]{Vphi_Au197.eps} &
  \includegraphics[scale=0.187]{Vphi_Pb208.eps} &
  \includegraphics[scale=0.187]{Gphi_He4.eps} \\
  \includegraphics[scale=0.187]{Gphi_C12.eps} &
  \includegraphics[scale=0.187]{Gphi_O16.eps} &
  \includegraphics[scale=0.187]{Gphi_Au197.eps} \\
  \includegraphics[scale=0.187]{Gphi_Pb208.eps} \\
\end{tabular}}
\caption{\label{fig:phinuclpot} Real $U_{\phi}(r)$ and imaginary $W_{\phi}(r)$ 
parts of the $\phi$--nucleus potentials for various nuclei.}
\end{figure}
In \fig{fig:phinuclpot} we show the $\phi$ potentials for some selected nuclei. We note that the results for $^{197}\text{Au}$ nucleus are 
presented here for the first time. One can see that the depth of $U_{\phi}(r)$ is sensitive to $\Lambda_{K}$, but $W_{\phi}(r)$ is not. 
Using these complex potentials, we calculate the $\phi$ single-particle
energies and absorption widths for various nuclei, considering the situation
where the $\phi$ is produced at rest. Then, under this condition, solving 
the Proca equation becomes equivalent to solving the Klein-Gordon equation
\begin{equation}
\label{eqn:kge}
(-\nabla^{2} + \mu^{2} + 2\mu V(r))\phi(\vec{r})
= \mathcal{E}^{2}\phi(\vec{r}),
\end{equation}
where $\mu$ is the reduced mass of the system in vacuum, and $V (r)$ is 
given  by \eqn{eqn:Vcomplex}. 
The calculated bound state energies ($E$) and absorption widths
($\Gamma$)~\cite{Cobos-Martinez:2017woo}, related to the complex
eigenvalue $\mathcal{E}$ by $E= \operatorname{Re}\mathcal{E}-\mu$ and 
$\Gamma= -2\operatorname{Im}\mathcal{E}$, respectively, are given in \tab{tab:bsenergies}
with and without $W_{\phi}(r)$.
When $W_{\phi}(r)=0$ the $\phi$ is expected to form bound states with
all the nuclei studied (values in parenthesis). However, $E$ is dependent on
$\Lambda_{K}$, increasing with it. For $W_{\phi}(r)\ne 0$ the situation
changes considerably. Whether or not the bound states can be observed
experimentally is sensitive to  the value of $\Lambda_{K}$. However, for the
largest value of $\Lambda_{K}$, which yields the deepest potentials, the
$\phi$ is expected to form bound states with all the nuclei studied. 
However, since the so-called dispersive effect of the absorptive potential
is repulsive, the bound states disappear completely in some cases, even though
they were found when $W_{\phi}(r)=0$. This feature is obvious for the $^{4}$He
nucleus, making it especially relevant to the future experiments, planned at
J-PARC and JLab using light and medium-heavy
nuclei~\cite{JPARCE29Proposal,JLabphiProposal}. 

\begin{table}
    \centering
\begin{minipage}[t]{0.55\textwidth}
   \centering
\addtolength{\tabcolsep}{-4pt}
\renewcommand{\arraystretch}{0.2}
\tiny
\begin{tabular}{ll|rr|rr|rr} 
\hline \hline
& & \multicolumn{2}{c|}{$\Lambda_{K}=2000$} &
\multicolumn{2}{c}{$\Lambda_{K}=3000$} & 
\multicolumn{2}{|c}{$\Lambda_{K}=4000$}  \\
\hline
 & $n\ell$  & $E$ & $\Gamma/2$ & $E$ & $\Gamma/2$ & $E$ & $\Gamma/2$ \\
\hline
$^{4}_{\phi}\text{He}$ & 1s & n (-0.8) & n & n (-1.4) & n & -1.0 (-3.2) & 8.3 \\
\hline
$^{12}_{\phi}\text{C}$ & 1s & -2.1 (-4.2) & 10.6 & -6.4 (-7.7) & 11.1 & -9.8
(-10.7) & 11.2 \\
\hline
$^{16}_{\phi}\text{O}$ & 1s & -4.0 (-5.9) & 12.3 & -8.9 (-10.0) & 12.5 & -12.6
(-13.4) & 12.4 \\
& 1p & n (n) & n & n (n) & n & n (-1.5) & n \\
\hline
$^{197}_{\phi}\text{Au}$
& 1s & -14.6 (-15.0) & 16.9  & -20.5 (-20.8) & 16.1 & -25.0 (-25.2) & 15.5 \\
& 1p & -10.9 (-11.6) & 16.2 & -16.7 (-17.2) & 15.5 & -21.1 (-21.4) & 15.0 \\
& 1d & -6.4 (-7.5) & 15.2 & -12.0 (-12.7) & 14.8 & -16.3 (-16.7) & 14.4 \\
& 2s & -4.6 (-6.1) & 14.6 & -10.1 (-11.0) & 14.3 & -14.3 (-14.9) & 14.0 \\
& 2p & n (-1.3) & n & -3.9 (-5.3) & 13.0 & -7.9 (-8.8) & 12.9 \\
& 2d & n (n) & n  & n (n) & n & -1.1 (-2.7) & 11.4 \\
\hline
$^{208}_{\phi}\text{Pb}$ & 1s & -15.0 (-15.5) & 17.4 & -21.1 (-21.4) & 16.6 &
-25.8 (-26.0) & 16.0 \\
& 1p & -11.4 (-12.1) & 16.7 & -17.4 (-17.8) & 16.0 & -21.9  (-22.2) & 15.5 \\
& 1d & -6.9 (-8.1) & 15.7 & -12.7 (-13.4) & 15.2 & -17.1 (-17.6) & 14.8 \\
& 2s & -5.2 (-6.6) & 15.1 & -10.9 (-11.7) & 14.8 & -15.2 (-15.8) & 14.5 \\
& 2p & n (-1.9) & n & -4.8 (-6.1) & 13.5 & -8.9 (-9.8) & 13.4 \\
& 2d & n (n) & n & n (-0.7) & n & -2.2 (-3.7) & 11.9 \\
\hline \hline
\end{tabular}
\end{minipage}\quad
\begin{minipage}[t]{0.425\textwidth}
\centering
\addtolength{\tabcolsep}{-4pt}
\renewcommand{\arraystretch}{0.72}
\tiny
\begin{tabular}{ll|r|r|r}
  \hline \hline
  & & \multicolumn{3}{c}{Cutoff $\Lambda_{D}$} \\
  \hline
& $n\ell$ & 2000 & 4000 & 6000\\
\hline
& & $E$ & $E$ & $E$ \\
\hline
$^{4}_{J/\Psi}\text{He}$
& 1s & n & -0.70 & -5.52 \\
\hline
$^{12}_{J/\Psi}\text{C}$
& 1s & -0.53 & -4.47 & -11.28 \\
\hline
$^{16}_{J/\Psi}\text{O}$
& 1s & -1.03 & -5.73 & -13.12 \\
\hline
$^{197}_{J/\Psi}\text{Au}$
& 1s & -4.09 & -10.49 & -19.09 \\
& 1p & -2.98 & -9.18 & -17.64 \\
& 1d & -1.66 & -7.53 & -15.80 \\
& 2s & -1.23 & -6.87 & -15.00 \\
& 1f & -0.20 & -5.64 & -13.66 \\
\hline
$^{208}_{J/\Psi}\text{Pb}$
& 1s & -4.26 & -10.84 & -19.67 \\
& 1p & -3.16 & -9.53 & -18.23 \\
& 1d & -1.84 & -7.91 & -16.41 \\
& 2s & -1.41 & -7.26 & -15.64 \\
& 1f & -0.39 & -6.04 & -14.30 \\
& 2p & -0.05 & -5.11 & -13.18 \\
\hline \hline
\end{tabular}
\end{minipage}
\caption{\label{tab:bsenergies}$\Phi$- and $J/\Psi$-nuclear bound state
energies ($E$) and and absorption widths ($\Gamma$).
Units are in MeV.}
\end{table}

\section{Nuclear-bound $J/\Psi$}

Following the same procedure as in the $\phi$ meson case, here we update
results for the $J/\Psi$-nuclear bound states, considering only the lightest
intermediate state in the $J/\Psi$ self-energy, namely the $D\overline{D}$
loop.
In the original studies~\cite{JPsiBoundStates}, the $J/\Psi$ self-energy
intermediate states involved the $D$, $\overline{D}$, $D^{*}$, and
$\overline{D^{*}}$ mesons. However, it turned out that the $J/\Psi$
self-energy has larger contributions from the loops involving the $D^{*}$, 
and $\overline{D^{*}}$ mesons, which is unexpected; see Krein {\it et al} 
in Ref.~\cite{vectormesonsinnuclmatt} for details on the issues involved.
In \tab{tab:bsenergies}, right panel, we present our updated  results for 
the $J/\Psi$-nuclear bound states, adding also the $^{197}\text{Au}$
nucleus for the first time.
We note that we have set the strong interaction width of the $J/\Psi$ to 
zero~\cite{JPsiBoundStates}, and therefore the $J/\Psi$ potentials are real
for all nuclei.
From these results, we expect that the $J/\Psi$ meson will form nuclear 
bound states for nearly all the nuclei considered, but some cases for
$^{4}$He, and that the signal for the formation should be experimentally 
very clear, provided that the $J/\Psi$ meson is produced  in recoilless
kinematics.
Thus, it will be possible to search for the bound states in a $^{208}$Pb nucleus at JLab after the 12 GeV upgrade. 

\section{Summary}

We have presented results for the $\phi$- and $J/\Psi$-nuclear bound states,
where  the vector meson potentials in nuclei have been obtained in the local
density  approximation from the vector meson self-energy in nuclear matter. 
The in-medium $K$ and $D$ masses as well as the the nuclear density
distributions for all nuclei but $^{4}$He are obtained in the QMC model.
From our results, w expect that the $\phi$ and $J/\Psi$ vector mesosn should form bound states for all five nuclei studied, provided that these vector mesons are produced in (nearly) recoilless kinematics.

\end{document}